# A Network Decoupling Method for Voltage Stability Analysis Based on Holomorphic Embedding

Qiupin Lai, Chengxi Liu and Kai Sun

*Abstract*—**This paper proposes a network decoupling method based on Holomorphic Embedding (HE) for voltage stability analysis. Using the proposed HE method with a physical load scaling factor *s*, it develops a set of decoupled two-bus circuit channels between a target bus and the swing bus. Accordingly, a complex-valued virtual index, named $\sigma(s)$, is introduced in each channel to assess the voltage stability of prospective operating conditions with ensured accuracy. Then the stability margin at each bus can be analyzed by visualizing its respective $\sigma(s)$ index on the $\sigma$ plane with a unified boundary of voltage collapse. Moreover, benefitted by the embedding with physical loading, the trajectory of $\sigma(s)$ for each bus with the variation of operating conditions can be analytically given about the scaling factor *s*. The effectiveness of proposed network decoupling method is verified on the IEEE 14-bus power system.**

## I. Introduction

With the development of the societal economy and the augment of power energy demand, power systems trend to be operated under heavier load and closer to the stability limits. Hence, it is critical for power system operators to monitor and analyze the system stability and security quickly and accurately. The current power flow analysis in most power system software is mainly based on the iterative numerical methods, which may suffer from divergence or give unfeasible solutions. To address these issues, the holomorphic embedding (HE) method is proposed in [1] and has been applied on saddle-noddle bifurcation estimation [2], network reduction [3], online voltage stability assessment [4], remote voltage control [5], voltage collapse diagnosis [6], etc.

Based on the theory of HE, this paper proposes a network decoupling method for voltage stability analysis. The main contributions are summarized as follows:

1) Inspired by the analytic solution of two-bus system, bulk power systems are decoupled to a set of equivalent two-bus circuit channels with corresponding complex-valued indices, i.e. $\sigma(s)$. Consequently, the voltage stability of all buses is visualized on a complex plane with a unified boundary of voltage collapse.

2) The varying operating conditions are embedded physically in the network decoupling method, so the analytic trajectory of $\sigma(s)$ can be used to assess the voltage stability of a prospective operating condition and identify the weak bus.

## II. Proposed Network Decoupling Method

Rigorously speaking, only a two-bus system has its analytic solution of power flow [7]. The basic idea of the proposed method is to decouple a multi-bus system into a set of two-bus circuit channels between each bus and the swing bus.

### A. The Decoupled Two-Bus Circuit Channel

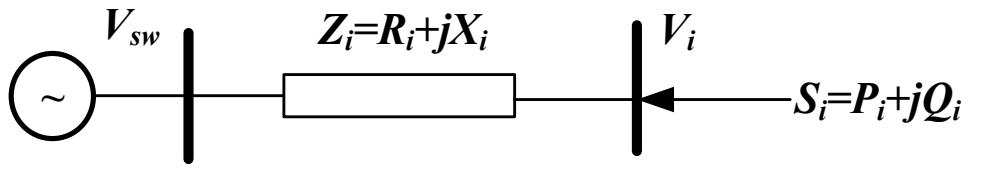

Fig. 1. Single-line-diagram of the decoupled two-bus channel for bus *i*.

The power flow equation (PFE) of the two-bus channel for bus *i*, shown in Fig. 1, can be obtained by Ohm's law.

$$\frac{V_i - V_{sw}}{Z_i} = \frac{S_i^*}{V_i^*} \tag{1}$$

It is more convenient to introduce a normalized voltage $U_i = V_i/V_{sw}$ and rearrange (1) as (2).

$$U_i = 1 + \frac{\sigma_i}{U_i^*} \tag{2}$$

where the complex-valued index $\sigma_i$, given by (3), is used to combine the parameters of virtual line impedance $Z_i$, swing bus voltage $V_{sw}$ and complex power injection $S_i$.

$$\sigma_i = \frac{Z_i S_i^*}{|V_{sw}|^2} \tag{3}$$

Since (2) is a quadratic equation, the solution is calculated directly by (4).

$$U_i = \frac{1}{2} \pm \sqrt{\frac{1}{4} + \sigma_{iR} - \sigma_{iI}^2} + j\sigma_{iI} \tag{4}$$

subject to (5):

$$\Delta = \frac{1}{4} + \sigma_{iR} - \sigma_{iI}^2 \geq 0 \tag{5}$$

where $\sigma_{iR}$ and $\sigma_{iI}$ denote the real part and imaginary part of $\sigma_i$, respectively. Note that (4) only has the upper solution with *plus* sign on its right-hand side [5]. The non-negative radicand in (5) is the necessary condition for existence of a feasible voltage solution, which forms the parabolic boundary on the complex plane of $\sigma$.

### B. Multi-Bus System

Inspired by the analytic solution of a two-bus channel as above, a multi-bus system composed of *m* PQ buses (sets of P ), *p* PV buses (sets of S ) and a swing bus, can be decoupled into $N$=$m$+$p$ equivalent two-bus circuit channels, as illustrated in Fig. 2. Through the proposed network decoupling method, each PQ bus or PV bus is individually on one channel connected with the swing bus $V_{sw}$. Every decoupled channel is denoted by its corresponding virtual index as an embedded holomorphic function $\sigma(s)$, able to represent the nonlinearity of PFEs. Once the coefficients of $\sigma(s)$ indices are solved, the bus voltages can be analyzed by plotting the $\sigma(s)$ of all buses on a unified complex plane with the parabolic boundary of voltage collapse.

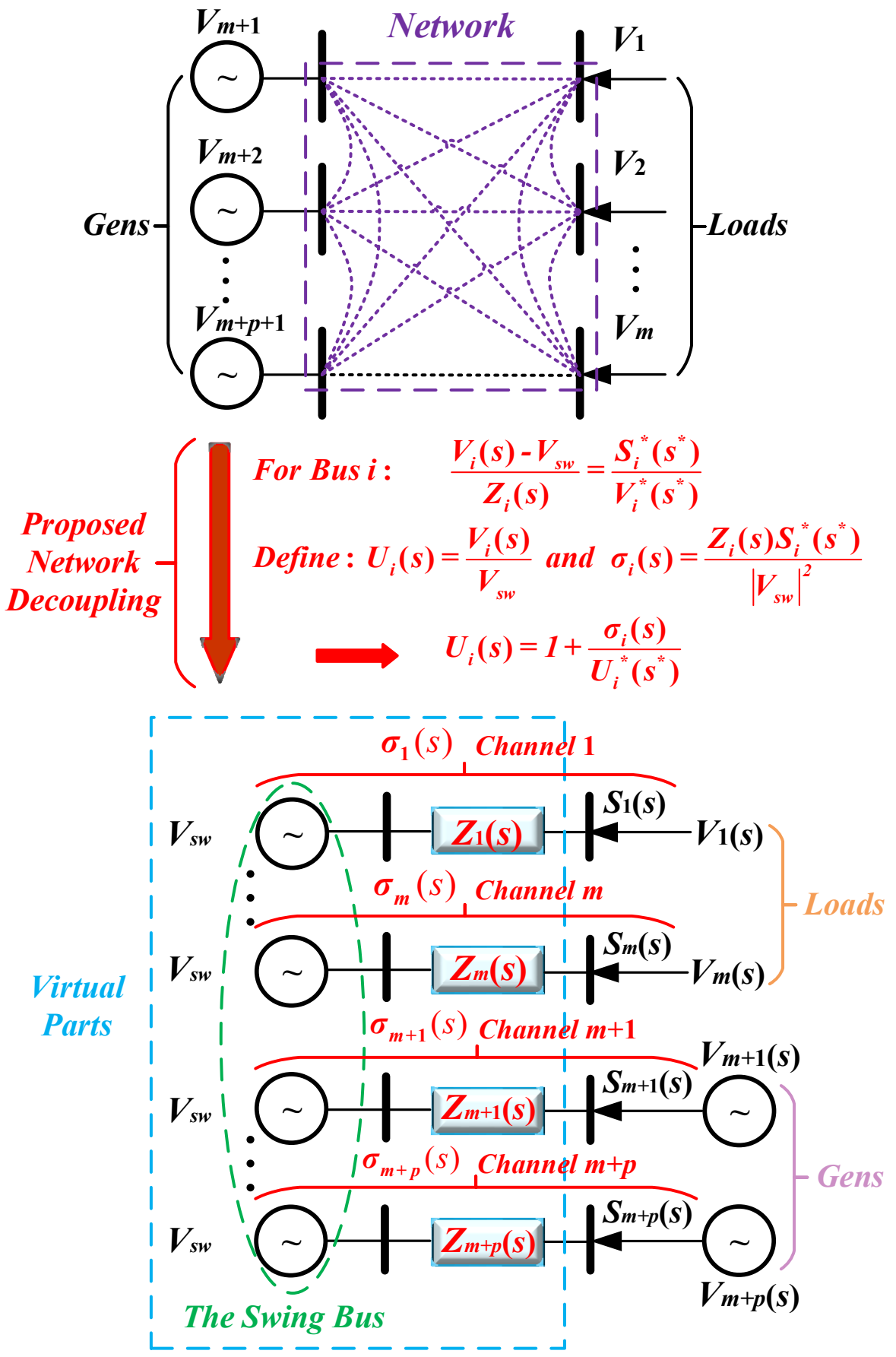

Fig. 2. Illustration of proposed network decoupling method for a multi-bus system with $m$ PQ buses, $p$ PV buses and a swing bus.

## III. Analytic Trajectory for Varying Conditions

### A. Embedding of a Physical Loading scale

Embedding a meaningful loading scale into PFEs endows the analytic trajectories with the applicability of varying operating conditions. For example, in (6), the loads and real power of generators are scaled uniformly by the factor $s$.

$$\begin{cases} V_{sw}[\sum\limits_{\substack{k=0\\k\in swing}}^{N} Y_{ik} + \sum\limits_{\substack{k=0\\k\in\mathcal{M}\cup\mathcal{P}}}^{N} Y_{ik}\left(1+V_{sw}M_k(s)\right)] = sS_i^*W_i^*(s^*), \forall i\in\mathcal{M} \\ V_{sw}[\sum\limits_{\substack{k=0\\k\in swing}}^{N} Y_{ik} + \sum\limits_{\substack{k=0\\k\in\mathcal{M}\cup\mathcal{P}}}^{N} Y_{ik}\left(1+V_{sw}M_k(s)\right)] = \left(sP_i - jQ_i(s)\right)W_i^*(s^*), \forall i\in\mathcal{P} \\ \left|V_{sw}\right|^2\left(1+V_{sw}M_i(s)\right)\left(1+V_{sw}M_i^*(s^*)\right) = \left|V_i^{sp}\right|^2, \forall i\in\mathcal{P} \end{cases} \quad (6)$$

where $V_{sw}$ is the voltage of swing bus, and $W_i$ is the reciprocal of $V_i$. For convenience, a new variable $M_i$ in (6) is introduced for calculation. The relation of $W_i$, $M_i$ and $\sigma_i$ is given by (7).

$$\begin{cases} M_i(s) = \sigma_i(s)W_i^*(s^*) \\ 1 = V_{sw}W_i(s) + \left|V_{sw}\right|^2 M_i(s)W_i(s) \end{cases}, \forall i\in\mathcal{M}\cup\mathcal{P} \quad (7)$$

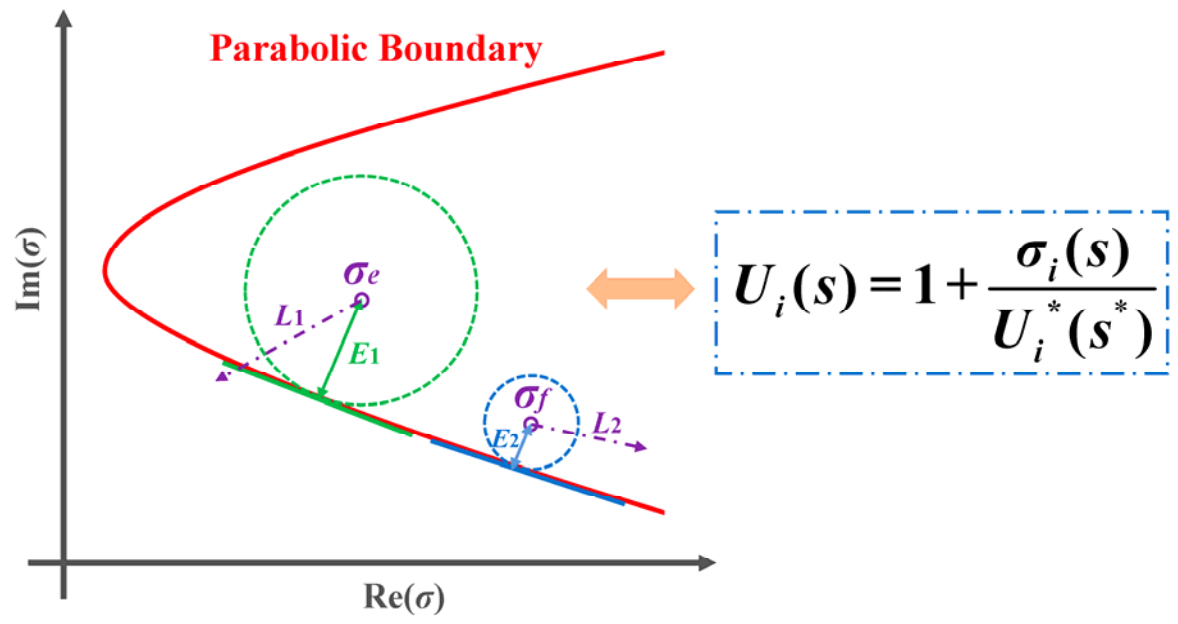

Fig. 3. Illustration of trajectories of $\sigma_i$ with varying operating conditions.

Please note that $U_i(s)=V_i(s)/V_{sw}$ is embedded as shown in Fig. 3 to match the practical meaning of PFEs in (6). The traditional scatter plots of $\sigma$ without trajectory can only assess the current condition for $s$=1 [6], [8]. It is inefficient for other operating conditions. Intuitively, the Euclidean distance to the boundary does not necessarily indicate the criticality of the associated bus. For instance, in Fig. 3, although the point of $\sigma_e$ seems farther from the boundary than that of $\sigma_f$ ($E1>E2$), $\sigma_e$ will reach the parabolic boundary with the trajectory of $L1$ before $\sigma_f$ with $L2$. Therefore, the trajectories of $\sigma(s)$, obtained by the proposed embedding method with practical meaning, can be applied for the voltage stability assessment of prospective operating conditions and identify the potential weak buses. The state-cum-trend trajectory ensures the effectiveness of $\sigma(s)$ index in any operating condition.

### B. Recursive Relation of Power Series Coefficients

In order to solve all the coefficients of power series, the recursive relations between a term and its previous terms are constructed by equating the coefficients of $s$ with same order at both sides of (6) and (7), and then separating equations into real and imaginary parts to obtain (8) in the Appendix, where all the *knowns* are moved to the right-hand side and the *unknowns* to the left-hand side.

Once the germ solutions of $W_i[0]$, $M_i[0]$ and $Q_i[0]$ are calculated by substituting $s$=0 into (6) and (7), all the following coefficients $M_i(s)$ and $W_i(s)$ are solved by (8) until the desired accuracy is achieved. The coefficients of $\sigma(s)$ are obtained by (9), which is derived by equating the coefficients of $s$ with same order at both sides of (7).

$$\sigma_i[n] = \frac{M_i[n] - \sum\limits_{\tau=0}^{n-1}\sigma_i[\tau]W_i^*[n-\tau]}{W_i^*[0]}, n\geq 0, \forall i\in\mathcal{M}\cup\mathcal{P} \quad (9)$$

## IV. Case Study on the IEEE 14-Bus Power System

As shown in Fig. 4, the proposed network decoupling method is verified on the IEEE 14-bus power system with the load conditions increasing successively from a light load condition (i.e. $s$ = 0.10 p.u.) to the critical collapse point. The corresponding $\sigma(s)$ indices of 13 buses except for the swing bus are collectively plotted on the complex plane of $\sigma$ in Fig. 5. The reactive power limit of each generator, which will result in discontinuity, is addressed by switching the violated PV bus to PQ bus, denoted by the purple circles in Fig. 5.

For the operating conditions with the same load increasing rate, the tested power system is operating closer and closer to the parabolic boundary of voltage collapse with the analytical trajectories of $\sigma(s)$ indices. The ranking of Euclidean distance from each $\sigma(s)$ index to the parabolic boundary continually changes because of their respective different trajectories. The analytic trajectory of $\sigma_{14}(s)$, i.e. the index of Bus 14, firstly arrives the parabolic boundary at $s$ = 1.75 p.u., indicating the weakest bus in the current load increasing condition, and the analytical expression of its trajectory is given as a piecewise function in Fig. 5. Due to the superiority of graphic visualization like Fig. 5, operators can quickly identify weak buses or vulnerable areas on the $\sigma$ plane in order to take a remedial voltage control action.

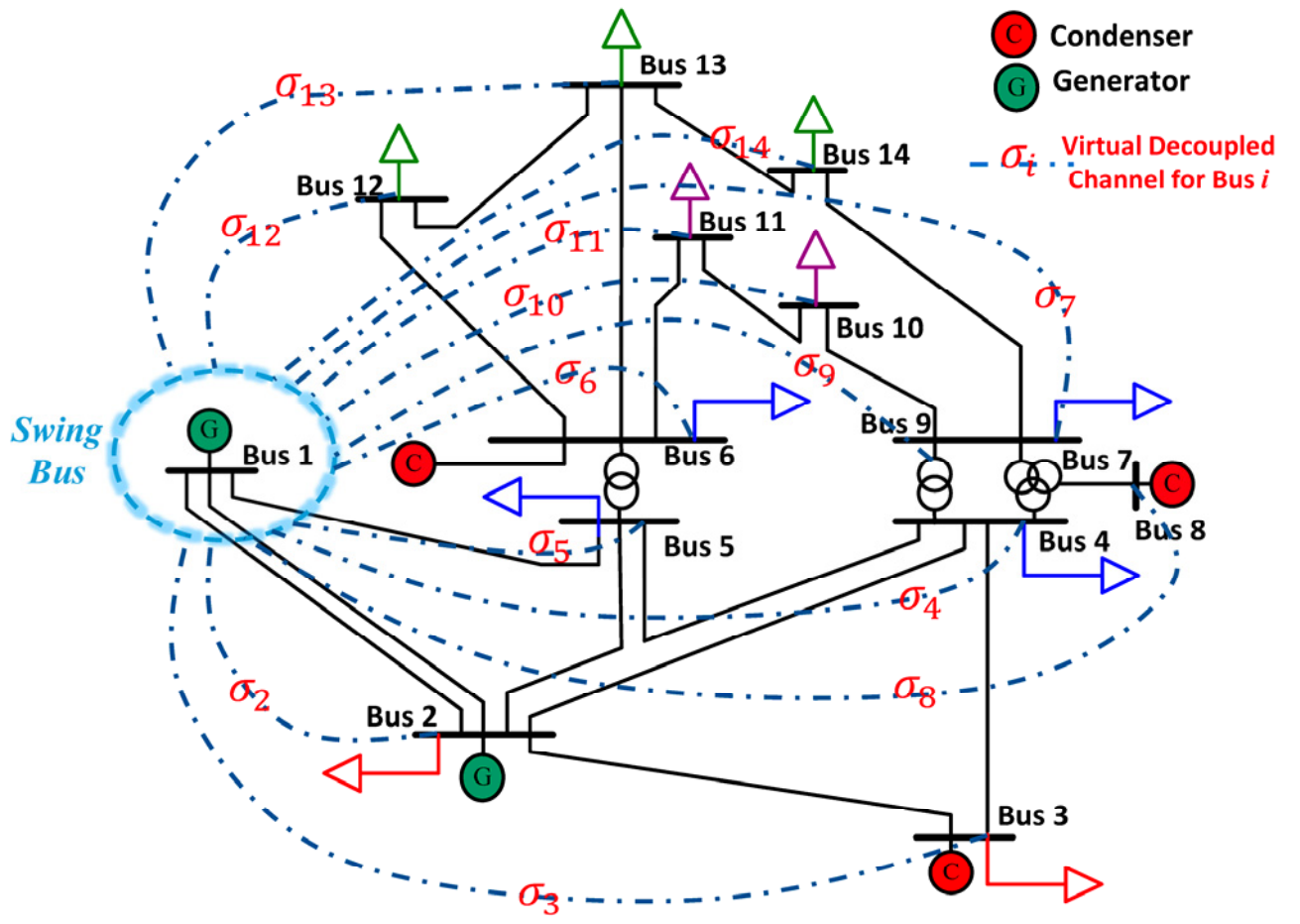

Fig. 4. The tested IEEE 14-bus power system with the swing bus of bus 1 and virtual decoupled channels of other buses.

## V. Conclusion and Discussion

This paper proposes a novel HE-based network decoupling method for voltage stability analysis. Through the proposed equivalent two-bus circuit channels with the swing bus, the visualization of $\sigma$ plots is convenient for stability analysis due to the unified parabolic boundary of voltage collapse. Besides, since the physical embedding method is applied to maintain the practical meaning with varying operating conditions, the analytical trajectories of $\sigma(s)$ indices are efficient for operators to evaluate voltage stability and avoid voltage collapse. Moreover, the nonlinearity of PFEs are also included in the $\sigma(s)$ indices to guarantee the accuracy of decoupling. How to timely and effectively dispatch the relevant loads and generations when the power system approaches emergent condition is the future work.

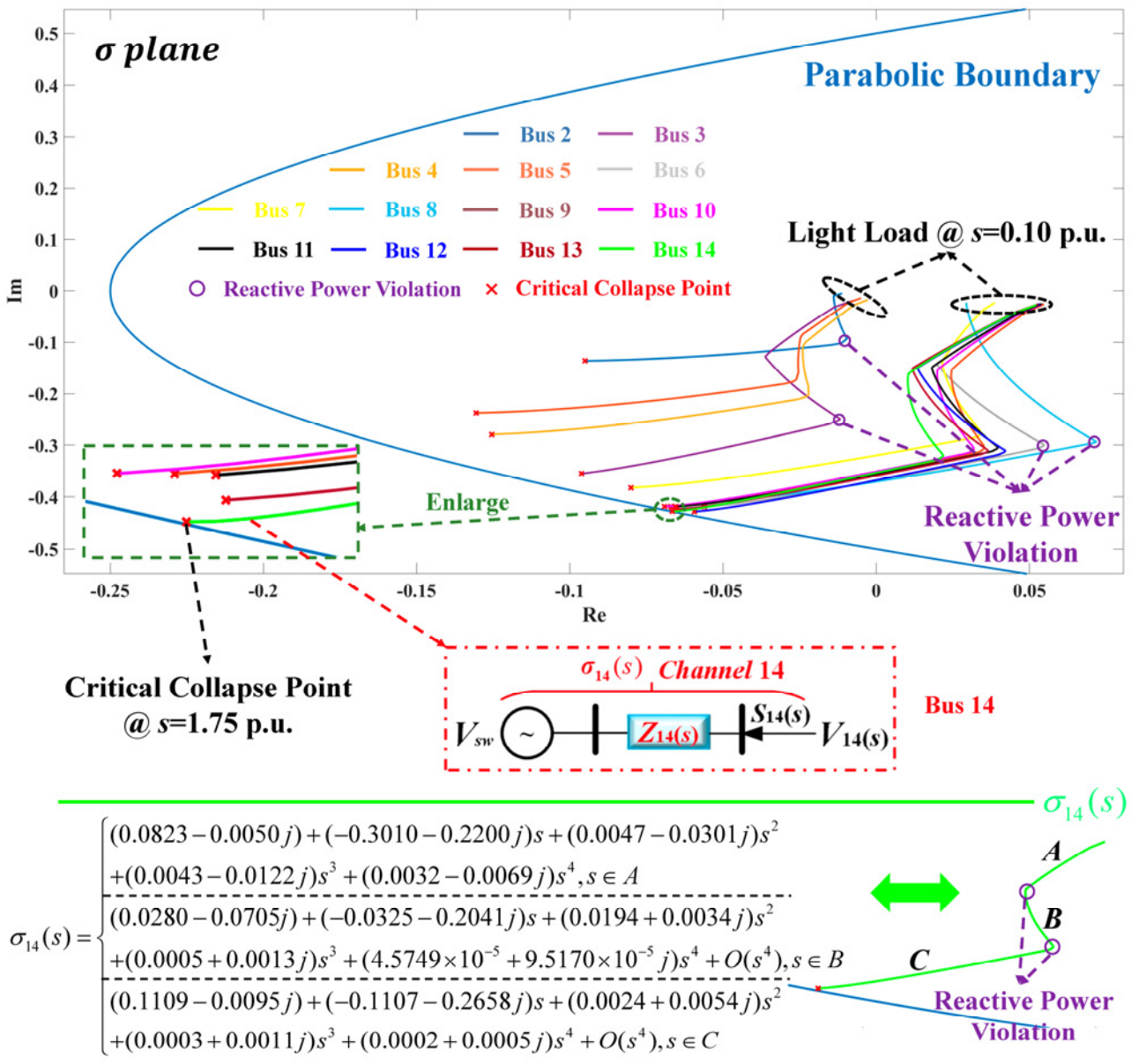

Fig. 5. The $\sigma$ plots with the load conditions increasing from light load condition to critical collapse point on the IEEE 14-bus power system.

## Appendix

$$
\begin{bmatrix}
G_{mimi} & -B_{mimi} & G_{mipi} & -B_{mipi} & & & \\
B_{mimi} & G_{mimi} & B_{mipi} & G_{mipi} & & & \\
G_{pimi} & -B_{pimi} & G_{pipi} & -B_{pipi} & W_{pi\Im}[0] & & Q_{pi}[0] \\
B_{pimi} & G_{pimi} & B_{pipi} & G_{pipi} & W_{pi\Re}[0] & Q_{pi}[0] & \\
 & & 2\frac{1+V_{sw}M_{pi\Re}[0]}{|V_{sw}|^2} & \frac{2M_{pi\Im}[0]}{V_{sw}} & & & \\
\frac{W_{mi}[0]}{V_{sw}} & & \frac{W_{pi}[0]}{V_{sw}} & & & 1+V_{sw}M_i[0] & \\
 & \frac{W_{mi}[0]}{V_{sw}} & & \frac{W_{pi}[0]}{V_{sw}} & & & 1+V_{sw}M_i[0]
\end{bmatrix}
\begin{bmatrix}
|V_{sw}|^2 M_{mi\Re}[n] \\
|V_{sw}|^2 M_{mi\Im}[n] \\
|V_{sw}|^2 M_{pi\Re}[n] \\
|V_{sw}|^2 M_{pi\Im}[n] \\
Q_{pi}[n] \\
W_{i\Re}[n] \\
W_{i\Im}[n]
\end{bmatrix}
=
\begin{bmatrix}
\Re\left(S_{mi}^{*}W_{mi}^{*}[n-1]\right) \\
\Im\left(S_{mi}^{*}W_{mi}^{*}[n-1]\right) \\
\Re\left(P_{pi}W_{pi}^{*}[n-1]-j(\sum_{\tau=1}^{n-1}Q_{pi}[\tau]W_{pi}^{*}[n-\tau])\right) \\
\Im\left(P_{pi}W_{pi}^{*}[n-1]-j(\sum_{\tau=1}^{n-1}Q_{pi}[\tau]W_{pi}^{*}[n-\tau])\right) \\
-V_{sw}\left(\sum_{\tau=1}^{n-1}M_{pi}[\tau]M_{pi}^{*}[n-\tau]\right) \\
-\Re\left(V_{sw}\sum_{\tau=1}^{n-1}M_{i}[\tau]W_{i}[n-\tau]\right) \\
-\Im\left(V_{sw}\sum_{\tau=1}^{n-1}M_{i}[\tau]W_{i}[n-\tau]\right)
\end{bmatrix}
, n \geq 1, \forall i \in \mathcal{M} \cup \mathcal{P}
\tag{8}
$$

PQ Bus

PV Bus

Eq. (6)

Eq. (7)